\documentclass[aps,prb,twocolumn,showpacs]{revtex4}
\usepackage{tabularx,graphicx}
\usepackage{epsfig}

\def\tg{$t_{2g}\/$\,} 
\def\eg{$e_{g}\/$\,}

\def\bil{La$_{2-2x}\/$Sr$_{1+2x}\/$Mn$_{2}$O$_{7}\/$\,}
\def\thb2{\frac{\theta_{i}}{2}}

\begin{document}

\author {T. Maitra$^{1}$\footnote{email: maitra@itp.uni-frankfurt.de}, 
A. Taraphder$^{2}$  and H.Beck$^{3}$\footnote{email: Hans.Beck@unine.ch} }   
\title { Charge order and phase segregation in overdoped bilayer manganites }
\affiliation{ $^{1}$Institut f\"ur Theoretische Physik, J. W. Goethe Universit\"at,
 Max-von-Laue-Strasse 1,
 60438 Frankfurt am Main, Germany \\ 
$^{2}$Department of Physics \& Meteorology and  Centre for 
Theoretical Studies,\\ Indian Institute of Technology, Kharagpur 721302 India  \\ 
$^{3}$Institute of Physics, University of Neuchatel, rue de Breguet 1, 
CH-2000 Switzerland }

\begin{abstract}

There have been recent reports of charge ordering around $x=0.5$ in the
bilayer manganites. At $x=0.5$, there appears to be a coexistence region
of layered A-type antiferromagnetc and charge order. There are also reports
of orbital order in this region without any Jahn-Teller effect. Based on 
physical grounds, this region is investigated from a model that incorporates 
the two \eg orbitals at each Mn site and a near-neighbour Coulomb repulsion. 
It is shown that there indeed is both charge 
and orbital order close to the half-doped region coincident with a layered 
magnetic structure. Although the orbital order is known to drive the magnetic 
order, the layered magnetic structure is also favoured in this system by 
the lack of coherent transport across the planes and the reduced 
dimensionality of the lattice. The anisotropic hopping across the \eg orbitals 
and the underlying layered structure largely determine the 
orbital arrangements in this region, while the charge order is primarily
due to the long range interactions.  

\end{abstract}


\maketitle 
\vspace{.5cm} 

\section{Introduction}

Bilayer manganites such as \bil, the $n=2$ member of the Ruddelsden-Popper 
series $(R,A)_{n+1}Mn_{n}O_{3n+1}$ 
(where $R$ and $A$ are rare-earth and alkaline-earth ions respectively), 
show \cite{ling} a distinct lack of symmetry in the nature of their 
ground states across $x=0.5$ just as their 3D counterparts. These systems 
have started drawing attention after colossal magnetoresistance (CMR) has 
been reported in them \cite{moritomo}. 
These layered systems not only show large magnetoresistance and a 
sequence of magnetic phases \cite{kimura,hirota}, they are very rich
in their charge, magnetic and orbital structures. 
Most of the studies in the bilayer system so far focussed on the Mn$^{3+}$
richer region of $x < 0.5$. The Mn$^{4+}-$rich region ($x > 0.5$) is 
now explored carefully both experimentally \cite{ling,billinge,cold} and 
theoretically \cite{tmat3,maezono} and a succession of magnetic phases 
A$\rightarrow$ C$\rightarrow$ G has been confirmed. The situation very 
close to and around $x=0.5$ is somewhat poorly understood yet. It emerged 
recently that there is a coexistence between charge ordered and layered 
A-type spin ordered state \cite{kubota,cold} there, giving way to the C-type
(or its polytype \cite{ling}) magnetic order at larger Mn$^{4+}$ concentration.
Both A-type (ferromagnetic layers coupled antiferromagnetically) and C-type 
(ferromagnetic chains coupled antiferromagnetically) phases have been found to be 
orbitally ordered. 

There are strong indications from several experimental groups that at and 
around $x=0.5$, a charge ordered state coexists with an A-type antiferromagnetic 
phase. Both neutron scattering \cite{argy,sury} and muon spin rotation corroborate 
this view.
A discontinuity in the muon precession frequency at $x=0.52$ has been
observed by Coldea et al.\cite{cold}. Recently Wilkins et al.,\cite{alta}
have reported that although there is clear evidence of both charge and
orbital order in \bil at $x=0.5$, there is no detectable signature for any 
Jahn-Teller (JT) distortion in their resonant X-ray diffraction study. 
JT distortion
of the MnO$_6$ octahedra begin to appear beyond $x=0.55$ \cite{billinge}. 
There is, though, a reasonably large (tetragonal) static distortion - 
the crystallographic $c/a$ ratio reaches a minimum \cite{ling} around 
the same composition, enhancing the AF super-exchange along $c$ direction. 
This is expected to favour the A-type spin order in that region. Unlike in 
the 3D manganites, there is no significant buckling of the bonds 
during distortion.  

It is now generally believed that the various magnetic structures owe their
origin to a large degree to the underlying orbital order \cite{akimoto,tmat3}.  
Models have been proposed \cite{khom,tmat1} 
for the manganites that incorporate the $e_g$ orbitals and the anisotropic 
hopping between them. The use of such models to the bilayer manganites 
(like \bil) has only had limited success \cite{maezono} though, the
A-phase is overestimated owing to the low dimensionality of the lattice. 
A much improved magnetic and orbital phase diagram was later obtained
\cite{tmat3} from a model that included the JT distortion as well as 
the proper choice of orbital degrees of freedom. 
The A-type AFM instability is indeed quite strong in the layered 
system \cite{ling}, extending from $x=0.42$ to 0.66 and at low temperatures.
The FM region of the phase diagram of some of the 3D manganites around $x=0.5$ 
is absent here, taken over by the A-type AFM phase. Having only two layers
separated by large distance from the adjoining bilayer in the c-direction
impedes a long range charge transport in that direction leading to stronger 
AF correlations along c-direction thereby contributing to an enhanced 
A-type instability. 
Beyond $x=0.66$, it is replaced by the C-type (and its polytype) 
order. The A- and C-phases are orbitally ordered and there is a connection 
between the preferred orbital order and the observed magnetic order. 

However, the presence of charge order and its relation to the underlying 
magnetic phases have not been investigated in any of the previous theoretical 
studies \cite{maezono,tmat3} in bilayer systems. 
The question of charge order and possible phase separation in the overdoped 
bilayer systems, therefore, still remains an open issue with increasing evidence 
in favour of such coexistence coming from the experiments. We attempt to address 
it in the following, starting from a model that incorporates the essential physical 
attributes of this region.  

\section{The model}

A quite general model for the bilayer manganites has been used \cite{tmat3}
to delineate the different orbital and magnetic structures for $x > 0.5$.  
It incorporates the degenerate \eg manifold and the physics of double exchange 
(DE) along with electron-electron interactions. This model can be adapted to 
investigate the charge order by including a longer range Coulomb term in the 
interaction part. Such a term has been known \cite{tmat2,jack} to give rise
to coexisting charge ordered state in the 3D manganites at $x=0.5$.  

\begin{widetext}

\begin{equation}
H = J_{AF} \sum_{<ij>} {\bf S_{i}}.{\bf S_{j}} - J_{H} \sum_{i} {\bf S_{i}}.
{\bf s_{i}} - \sum_{<ij>\sigma,\alpha,\beta}t_{i,j}^{\alpha \beta} 
c_{i,\alpha,\sigma}^{\dagger} c_{j,\beta,\sigma} + H_{int} 
\end{equation}
\end{widetext}

As usual the charge and spin dynamics of the conventional DE model\cite{zener} 
operate here too, with additional degrees of freedom coming from the degenerate 
\eg orbitals ($\alpha, \beta$ stand for the \eg orbitals) and hopping across them, 
which is determined by their symmetry. In the reasonably large Hund's coupling 
limit, which is relevant for the studies on manganites, the DE mechanism implies 
that an electron can hop onto a site if core spins at that site (as well as the spin
of any \eg electron there) are parallel to its spin. Mobility reduces drastically 
with increasing J$_H$ if they are antiparallel.  
In Eqn.(1) ${\bf S_{i}}$ and ${\bf s_i}$ are the \tg and \eg spins at 
site $i$ and $J_H$
and $J_{AF}$ are the Hund and super-exchange (SE) coupling respectively. 
$H_{int}= U^{\prime} \sum_{i\sigma\sigma^{\prime}}\hat{n}_{i1\sigma}
\hat{n}_{i2\sigma^{\prime}}+V \sum_{i}\hat{n}_{i}\hat{n}_{j}$    
contains the on-site inter-orbital and the near-neighbour 
Coulomb repulsion term. The intra-orbital term can be ignored as $J_H$ in 
manganites is fairly large, preventing double occupancy in any given 
Mn 3d orbital\cite{hotta}. The exchange interaction between two bilayers 
is known to be at least a 100 times weaker \cite{fujioka} than the 
intra-bilayer exchange. Two bilayers are also well-separated in an unit cell 
and intervened by the rare-earth ions. One can, therefore, use only one 
bilayer for a reasonable description of the system\cite{maezono}.  
In general, in manganites, there is a strong JT coupling and that is
included in the Hamiltonian. As we mentioned above, in the bilayer systems
(e.g. in \bil), the JT coupling is weak and does not play a major role 
\cite{billinge,alta} in the region $x < 0.6$.  
In the absence of electron-lattice coupling, the kinetic energy (KE) of 
electrons in the $e_g$ band and the Hund's coupling between $t_{2g}$ and 
$e_g$ spins compete with the antiferromagnetic SE interaction leading to 
a variety of magnetic and orbital structures. Considering that there are two 
$e_g$ orbitals, the nominal band filling is $\frac{1-x}{4}$. 

Typical values of the interaction and band parameters for the bilayer systems 
are in the same range as in the 3D manganites. As in their 3D counterparts, 
the Hund coupling and Coulomb correlations are the largest scale of energy 
\cite{hotta} in the problem. We neglect the inter-orbital Coulomb term in 
the following discussion of charge order. The inter-orbital Coulomb term
does not have strong effect in the doping range $x$ less than or around
0.5 primarily because of the low filling ($(1-x)/4$) making two orbitals
at the same site less likely to be occupied simultaneously. As we 
see below, the magnetic order also prefers preferential occupancies in one of 
the two \eg orbitals in both A- and C-phases and the $U^\prime$ term only 
enhances that (discussed later).  

Treating the 
t$_{2g}$ spins semiclassically \cite{zener}, the SE contribution to the 
ground state energy 
becomes $E_{SE}=\frac {J_{AF}S_0^2}{2}(2cos\theta_{xy}+cos\theta_z)$ where 
$\theta_{xy}$ and $\theta_{z}$ are the angle between the near-neighbour 
\tg spins in the $xy$ plane and $z$ direction respectively. In the Ferromagnetic
state, $\theta_{xy}=\theta_z=0$ while in the A-phase $\theta_{xy}=0$ and $\theta_z
=\pi.$ 

In the limit of infinite J$_H$, the \eg electron quantization axes at each site $i$ 
is rotated in the local coordinate frame to make it parallel to ${\bf S}_i$. 
This is accomplished quite easily by the spin-1/2 rotation matrix 
$exp(i \frac{\phi_{i}}{2}\sigma_{z})\, exp(i \frac{\theta_{i}}{2}\sigma_{y})
\,exp(-i\frac{\phi_{i}}{2}\sigma_{z})$
operating on a two component spinor.  Allowing the core spins at each site to cant 
in the $xz-$plane and neglecting the phase term (the Berry phase) appearing from 
the transformation, it is straightfoward to show \cite{zener} that the effective 
hopping matrix elements are $t_{xy}=tcos(\theta_{xy}/2)$ and 
$t_z=tcos(\theta_z/2).$ In this level of approximation, the diagonalisation of 
the KE part of $H$ reduces to solving the $2\times 2$ matrix equation
$|| \epsilon_{\alpha\beta}-\omega\delta_{\alpha\beta} || =0$ for a system of
spinless fermions. The matrix elements are obtained from the standard table of 
the overlap integrals \cite{slater} involving $d_{x^2-y^2}$ and $d_{3z^2-r^2}$ 
orbitals on neighbouring sites of a square lattice and the dispersions are 

$$\epsilon_{11}=-2t_{xy}(cosk_x+cosk_y)$$
$$\epsilon_{12}=t_{21}=-\frac{2}{\sqrt{3}}2t_{xy}(cosk_x-cosk_y)$$
$$\epsilon_{22}=-\frac{2}{3}t_{xy}(cosk_x+cosk_y)-\frac{8}{3}t_zcosk_z.$$

Using the form of $t_{xy}$ and $t_z$ in terms of $\theta_{xy}$ and $\theta_z$ 
in the J$_H\rightarrow \infty $ approximation, the two energy bands are obtained
by the diagonalization of the above matrix.  
In the uncanted A- and C-phases, the dispersions for A- and C-phases become 
two- and one-dimensional respectively. In the pure FM phase, the DOS is 
three dimensional. However, 
in the bilayer system, owing to the absence of dispersion in the $z$ direction,
the DOS in both FM and A phases show two dimensional character as shown in
Fig. 1. There being only two $k-$points in the $z-$direction, the DOS for 
the C-phase has two delta functions centred at $\pm \frac{8t_z}{3}$. 
In all the calculations that follow,  all  energies are measured in terms 
of the overlap between the  $d_{3z^{2}-r^2}$ orbitals along $z-$direction 
$t^{\hat{z}}=t$, which has a typical value about 0.25eV in manganites.

\begin{figure}[!]
  \begin{center}
       \epsfig{file=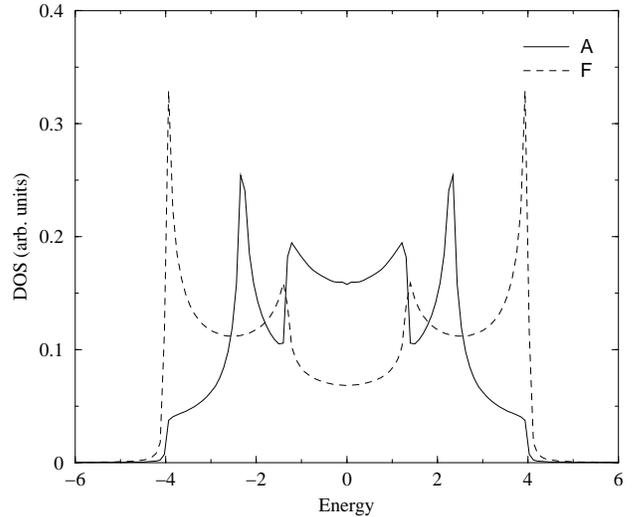,height=7cm}
    \caption{ Density of states in pure FM and A phase. } 
    \label{figure 1}
\end{center}
\end{figure}

\section{Charge and orbital ordering}

In order to look for charge ordering in the bilayer manganites, we treat
the nearest neighbour Coulomb interaction $V\sum_{<ij>}\hat{n}_i\hat{n}_j$ in 
the mean 
field approximation, $<\hat{n_i}>=n+C_0exp(iQ.r_i)$ where $C_0$ is the charge order 
parameter and $n$ is the average number of electrons per site. We take the
usual staggered ordering $Q=(\pi,\pi,\pi)$. The charge order parameter $C_0$ is
then calculated self-consistently. The mean-field approximation is known to 
work quite well \cite{tmat1,hotta,maezono} for the ground state properties 
in the manganites. In the infinite $J_H$ limit, the electronic part of the 
Hamiltonian is a $4\times 4$ matrix $\sum_{{\bf k},\alpha\beta} 
\epsilon_{{\bf k}{\alpha\beta}} \tilde{c}_{{\bf k}\alpha}^{\dagger} \tilde{c}_{
{\bf k}\beta}-\Delta \sum_{{\bf k},\alpha} \tilde{c}_{{\bf k}\alpha}^{\dagger}
\tilde{c}_{{\bf k+Q}\alpha}$ (where $\tilde{c}$ represents the locally rotated \eg 
electron operators described earlier) and this is diagonalized at each of the $\bf k$
points on a momentum grid. 


The bands obtained thereof are filled upto a chemical potential and the order 
parameter $\Delta=zVC_0$ ($z$ is the number of nearest neighbour) calculated along 
with the filling. The process
is repeated until self-consistency is achieved as is customary in the mean-field
theory. Charge order is indeed observed in a region $x\ge0.5$ when $V$ reaches 
a critical value, similar to the 3D manganites\cite{tmat2,jack}. But unlike
in the 3D case, we do not observe any F-phase coexisting with the CO in this
region. As reported in previous work \cite{maezono,okamoto} A-phase instability 
is quite strong in the layered manganites owing to the 2D structure of its DOS 
and even for $V=0$, there is no F-phase for $x \ge 0.5.$\cite{tmat3}. The
A-CO coexistence region extends in a region above $x=0.5$ (discussed later). 
In this limit of infinite $J_H$, the value of $V$ for which the  
CO phase appears depends strongly on the canting angle $\theta_z$ away
from the A-phase (towards an F-phase). In Fig. 2 is shown the the dependence
of the CO order parameter on $V$ at three different angles ($\theta_z
=180\deg$ being the pure A-phase). The inset shows how $C_0$ varies with
canting at a typical $V=0.7.$ In the infinite Hund's coupling limit, the
only way an electron gains KE in and AF configuration of background spins
is via the canting of them\cite{deG}, thereby generating an effective non-zero
hopping across. This has been observed\cite{tmat1} in the 3D manganites
as well close to its G-C phase boundary (for $x\simeq 1$). As we observe 
here, a CO phase does not coexist with an F phase (which is energetically
unfavourable to an A-phase in this region) and therefore the
increased canting requires larger $V$ to bring about the charge ordering. 
At a fixed $V$, therefore, the CO order parameter reduces with increased
canting away from A-phase.  

\begin{figure}[!]
  \begin{center}
       \epsfig{file=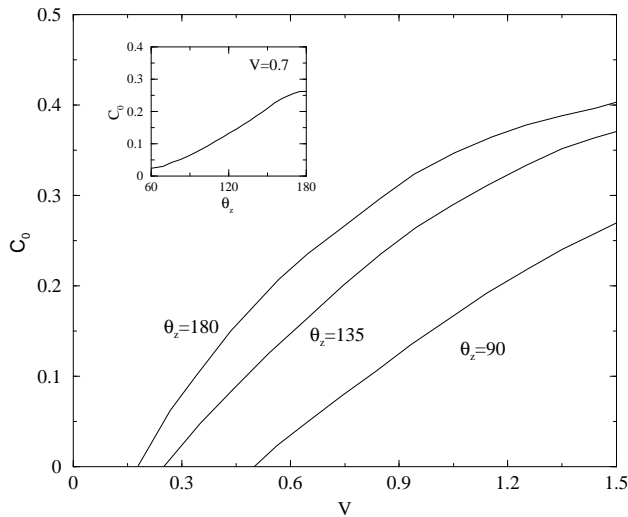,height=7cm}
    \caption{ Charge order parameter as a function of $V$ for three different
angles $\theta_z$ in the limit $J_{H}\rightarrow \infty$. In the inset is shown 
the variation of charge order parameter with $\theta_z$ for a fixed $V=0.7$. }  
    \label{figure 2}
\end{center}
\end{figure}

The appropriate limit for any of the manganites is, of course, a large
but finite value of the Hund's coupling as is generally believed. In this limit, 
we can treat the
core spins semi-classically again and the spin degrees of freedom for the $e_g$ 
electrons are reintroduced. In order to take care of this, the second term
in the Hamiltonian (1) would have to be treated now in various possible
ground states.   
We choose ${\bf S}_i={\bf S}_{0}\exp (i{\bf q.r_i})$  to represent a homogeneous 
spin configuration, where ${\bf q}$ determines different spin arrangements for 
the $t_{2g}$ spins \cite{tmat2}. The second term in eqn.(1) becomes
$-J_HS_0 \sum_{{\bf k}\alpha\sigma} \sigma c_{{\bf k}\alpha\sigma}^{\dagger}
c_{{\bf k+q}\alpha\sigma}.$ The near-neighbour Coulomb
term is treated in the mean-field as above. In this semi-classical 
approximation for the $t_{2g}$ spins the Hamiltonian (1) reduces, then, 
to a $16\times 16$ matrix\cite{tmat2} at each $\bf k$ point.  

This mean-field Hamiltonian is again diagonalised at each {\bf k}-point 
on the momentum grid. The ground state energy is calculated for different 
magnetic structures. The CO order parameter is also determined self-consistently. 
Four different magnetic structures are relevant for the experimental 
phase diagram (with q values in the parentheses), A-type ($0,0,\pi$), C-type 
($\pi,\pi,0$) - the usual C-phase with FM chains along c-direction, 
and the 3D AFM G-type ($\pi,\pi,\pi$). There is also the C$^\prime$-type
polytype structure \cite{ling,tmat3}, with ${\bf q}= (\pi,0,\pi)$ 
This is same as a C-type, only that its FM ordering is along y-direction as
reported by Ling et al. The magnetic structure with minimum 
ground state energy and the CO order parameter are determined for each set of 
parameters ($x$, $J_H$, $J_{AF}$) for the range of doping ($0.5 < x \le 1$).  

\begin{figure}[!]
  \begin{center}
       \epsfig{file=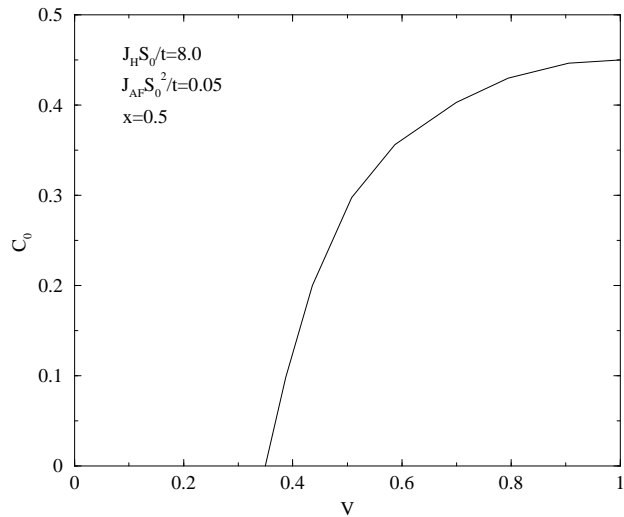,height=7cm}
    \caption{ Charge order parameter as a function of $V$ at $x=0.5$.  } 
    \label{figure 3}
\end{center}
\end{figure}

\begin{figure}[!]
  \begin{center}
       \epsfig{file=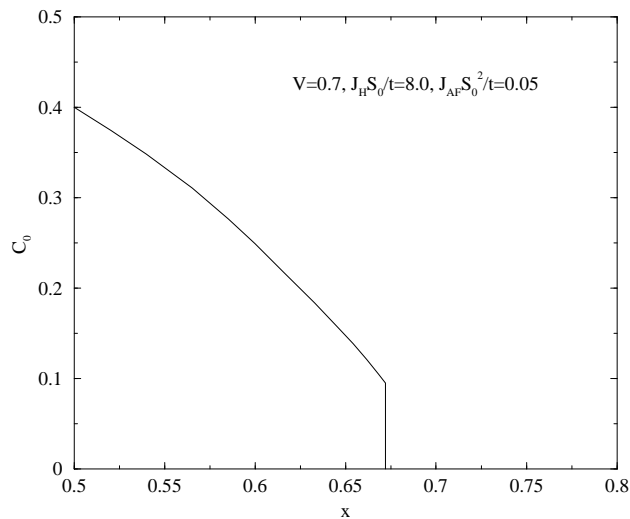,height=7cm}
    \caption{ Variation of charge order as $x$ is changed from 0.5 for a 
fixed $V=0.7$. The jump in the order parameter indicates the A-CO to A boundary} 
    \label{figure 4}
\end{center}
\end{figure}

Charge order is observed in the region $ 0.67 > x \ge 0.5$. Fig. 3 shows the
CO order parameter as a function of $V$ for $x=0.5$. The underlying spin 
order in this region is found to be A-type and there is a A-CO to A transition  
at around $x=0.67$ for $V=0.7.$ There is a jump in the CO order parameter as
a function of $x$, shown in Fig. 4, indicating a first order transition.
The coexistence region of CO and A is a manifestation of this first order
transition. Based on the observed variation of these different charge and
spin ordered states, we obtained a phase diagram in the $x-J_H$ plane for
$V=0.7$.

\begin{figure}[!]
  \begin{center}
       \epsfig{file=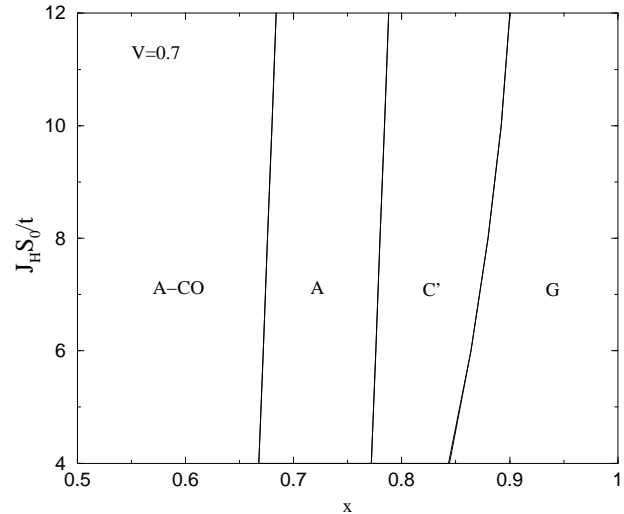,height=7cm}
    \caption{ Phase diagram, in the $J_{H}S_{0}$ versus filling $x$ plane. The terms
A, C$^\prime$, G, CO etc. are explained in the text.} 
    \label{figure 5}
\end{center}
\end{figure}

The observed phase diagram, shown in Fig. 5, is markedly similar 
to the one obtained
by Ling and coworkers. In addition it also has the charge ordered phase
coexisting with the A-phase in the region close to $x=0.5$. As reported by 
Coldea et al.,\cite{cold}, there is indeed a region of A-CO coexistence
between $x=0.5$ and $x=0.65$. We note that we have also tried a charge order
with wave vector ${\bf Q}=(\pi,\pi,0)$ instead of the fully staggered one
and observed that energetically it is very close to the staggered one. This is 
primarily due to the absence of coherent charge transport in the c-direction
making a charge uniform state in that direction nearly degenerate with a charge 
ordered one. We note in passing that we did not observe the well known CE 
phase\cite{kanamori} in our calculation even at $x=0.5$. This is similar
to the findings in the $V=0$ case studied earlier\cite{maezono,tmat3}.  
The region of no long range order just above the A-CO coexistence reported by
both Ling et al. and Coldea et al. is beyond the scope of the treatments
here. It is at this point useful to note that a region of two phase coexistence 
is not very stable against long range Coulomb interactions and in real 
systems one would probably observe a microscopically phase separated mixture of 
one phase in another. The region of no LRO could well be such a region, around 
the A-CO to A transition and with the competing C$^\prime$ phase energetically very 
close by.  

\begin{figure}[!]
  \begin{center}
       \epsfig{file=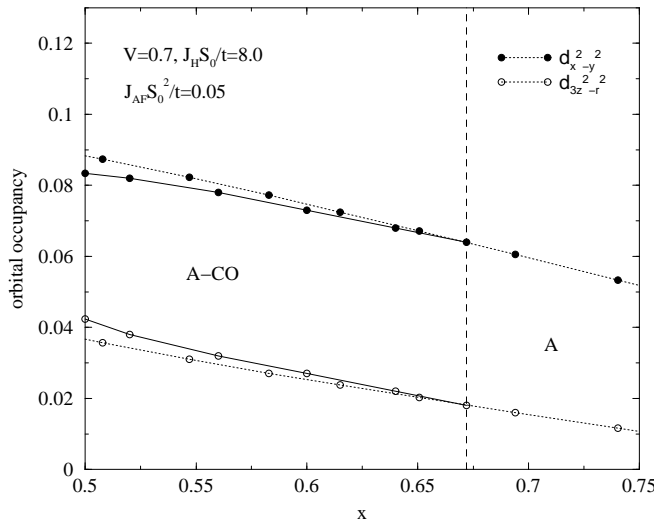,height=7cm}
    \caption{ Orbital occupancies in the A-CO and A-phases. The upper curves
are for $d_{3z^{2}-r^2}$ while the lower ones are for the $d_{x^{2}-y^{2}}$ 
orbitals. 
The dotted lines are for comparison with $V=0$ situation where no CO is present. } 
    \label{figure 6}
\end{center}
\end{figure}

In the absence of CO the phases A and C$^\prime$ are both orbitally ordered
\cite{tmat3} and we look for this in the presence of the near-neighbour Coulomb 
term also. We do indeed find very similar orbital occupancies here too. There
is a predominant occupancy in the $d_{3z^{2}-r^2}$ orbital over the $d_{x^{2}-y^{2}}$
orbital indicating orbital ordering. The occupancies in the two orbitals must add up
to the total electron density ($\frac{1-x}{4}$). 
The effect of charge ordering is seen to be small, shown in Fig. 6, reducing the 
orbital order (measured by the imbalance in the orbital occupancies) in the 
A-CO phase from its value in the absence of CO. There is no
effect in orbital occupancies at the A-CO to A boundary, they smoothly continue
into the pure A-phase. The presence of orbital order here is not contingent
upon the underlying lattice distortion or JT effects. This conforms to the
recent observation of Wilkins et al.\cite{alta} that even without any JT distortion,
there is a pronounced orbital order. In the region of C$^\prime$-phase, the 
orbital order is identical (predominantly $d_{3y^{2}-r^2}$ over $d_{z^{2}-x^{2}}$) 
to the one reported 
earlier\cite{tmat3} for the undistorted bilayer system and are not shown here.  
We have looked into the effect of an inter-orbital repulsion term on the
phase diagram and charge ordering and except for a slight enhancement of the
orbital order, found it to have little effect on the phase diagram. The
intra-orbital Coulomb term is known to have very little effect\cite{hotta} in 
the mean-field theory in the region of large J$_H$ and we haven't considered 
it in the above. 

\section{conclusion}

Motivated by the observation of charge ordering in the region $x\ge 0.5$ in the
bilayer systems, we have investigated it from a model that has been quite
useful in understanding the overdoped manganites. The charge order has been
found in our calculation in the same region where it has been seen experimentally.
There is also a region of phase separation abutting the experimentally observed
no LRO region. We observe orbital order even in the absence of JT distortion as
reported recently in the same region of doping. It would be interesting to see more
experiments on the region of no LRO in the phase diagram. The possibility of canting
of spins is yet not ruled out and we have seen trends for it for very large Hund's
coupling in our calculations. 
\vspace{0.3cm} 

\noindent {\bf Acknowledgement} 
We acknowledge useful discussions with M. Capezzali and G. V. Pai.


\begin{thebibliography}{999}
\bibitem{ling} C. D. Ling et al., Phys. Rev. B {\bf 62}, 15096 (2000).
\bibitem{moritomo} Y. Moritomo et al., J. Phys. Soc. Jpn. {\bf 67}, 405 (1998).
\bibitem{kimura} T. Kimura et al., Phys. Rev. B {\bf 58}, 11081 (1998). 
\bibitem{hirota} K. Hirota et al., J. Phys. Soc. Jpn. {\bf 67}, 3380 (1998).
\bibitem{cold} A. Coldea et al., Phys. Rev. Lett. {\bf 89}, 277601 (2002).  
\bibitem{billinge} X. Qiu et al., cond-mat/0307652. 
\bibitem{tmat3} Tulika Maitra and A. Taraphder, Europhys. Lett. {\bf 65}, 262 
(2004). 
\bibitem{maezono} R. Maezono and N. Nagaosa, Phys. Rev. B. {\bf 61}, 1825 
(1998); R. Maezono, S. Ishihara and N. Nagaosa, Phys. Rev. B. {\bf 58}, 
11583 (1998).
\bibitem{kubota} M. Kubota et al., J. Phys. Chem. Solids {\bf 60}, 116 (1999);
M. Kubota et al., cond-mat/9902288.
\bibitem{argy} D. N. Argyriou et al., Phys. Rev. B. {\bf 61}, 15269 (2000).  
\bibitem{sury} T. Chatterjee et al. Phys. Rev. B. {\bf 61}, 570 (2000).  
\bibitem{alta} S. B. Wilkins et al., cond-mat/0412435.  
\bibitem{akimoto} T. Akimoto et. al, Phys. Rev. B {\bf 57}, R5594 (1998).  
\bibitem{khom} J. van den Brink and D. Khomskii, Phys. Rev. Lett {\bf 82}, 1016
(1999).
\bibitem{tmat1} Tulika Maitra and A. Taraphder, Europhys. Lett. {\bf 59}, 896 
(2002). 
\bibitem{hotta} T. Hotta, A. Malvezzi and  E. Dagotto, Phys. Rev. B {\bf 62},
9432 (2000); E. Dagotto, T. Hotta and A. Moreo, Phys. Reports {\bf 344}, 1 
(2001). 
\bibitem{fujioka} H. Fujioka et al., cond-mat/9902253; K. Hirota et al.,
cond-mat/0104535. 
\bibitem{zener} C. Zener, Phys. Rev. {\bf 82}, 403 (1951); P. W. Anderson and
H. Hasegawa, $ibid$  {\bf 100}, 675 (1955). 
\bibitem{slater} J. C. Slater and G. F. Koster, Phys. Rev. {\bf 94}, 1498
(1954).
\bibitem{tmat2} Tulika Maitra and A. Taraphder, Phys. Rev. B, {\bf 66}, 174416
(2003). 
\bibitem{jack} G. Jackeli, N. B. Perkins and N. M. Plakida, Phys. Rev. B 
{\bf 62}, 372 (2001).
\bibitem{okamoto} S. Okamoto, S. Ishihara and S. Maekawa, Phys. Rev. B 
{\bf 63}, 104401 (2001). 
\bibitem{deG} P. G. de Gennes,  Phys. Rev. {\bf 118}, 141 (1960).  
\bibitem{kanamori} K. Kanamori, J. Appl. Phys. {\bf 14}S (1960) ; J. B. Goodenough,
Phys. Rev. {\bf 100}, 564 (1955). 
\end{thebibliography}
\end{document}